\documentclass[sigconf]{acmart}

\usepackage{subcaption}
\usepackage[capitalise]{cleveref}
\usepackage{tabularx}
\usepackage{array,makecell}
\usepackage{enumitem}
\usepackage{siunitx}
\usepackage[most]{tcolorbox}

\newcommand{\figvspace}{\vspace{0.5em}}%

\newcommand{\RQ}[1]{RQ\,#1}

\newtcolorbox{resultsbox}[2][]{enhanced,boxsep=0pt,top=\belowrulesep,bottom=\aboverulesep,left=0pt,right=0pt,boxrule=0pt,arc=0pt,colframe=white, 
	colback=white,
	coltitle=black,
	title={#2},
	detach title,
	before upper={\strut\makebox[\leftmargin][l]{\textbf{\tcbtitle}}},
	after upper={\strut},
	borderline north={-\heavyrulewidth}{0pt}{black, line width=\heavyrulewidth},
	borderline south={-\heavyrulewidth}{0pt}{black,line width=\heavyrulewidth}
	,#1}

\newcolumntype{L}{>{\raggedright\arraybackslash}X}
\newcolumntype{R}{>{\raggedleft\arraybackslash}X}

\copyrightyear{2025}
\acmYear{2025}
\setcopyright{cc}
\setcctype{by}
\acmConference[FSE Companion '25]{33rd ACM International Conference on the
Foundations of Software Engineering}{June 23--28, 2025}{Trondheim, Norway}
\acmBooktitle{33rd ACM International Conference on the Foundations of
Software Engineering (FSE Companion '25), June 23--28, 2025, Trondheim,
Norway}\acmDOI{10.1145/3696630.3728563}
\acmISBN{979-8-4007-1276-0/2025/06}

\begin{document}

\title{When Should I Run My Application Benchmark?}
\subtitle{Studying Cloud Performance Variability for the Case of Stream Processing Applications}

\author{Sören Henning}
\orcid{0000-0001-6912-2549}
\affiliation{%
  \institution{Dynatrace Research}%
  \city{Linz}%
  \country{Austria}%
}
\email{soeren.henning@dynatrace.com}

\author{Adriano Vogel}
\orcid{0000-0003-3299-2641}
\affiliation{%
    \institution{Dynatrace Research}%
    \city{Linz}%
    \country{Austria}%
}
\email{adriano.vogel@dynatrace.com}

\author{Esteban Perez-Wohlfeil}
\orcid{0000-0002-4415-6694}
\affiliation{%
    \institution{Dynatrace Research}%
    \city{Linz}%
    \country{Austria}%
}
\email{esteban.wohlfeil@dynatrace.com}

\author{Otmar Ertl}
\orcid{0000-0001-7322-6332}
\affiliation{%
    \institution{Dynatrace Research}%
    \city{Linz}%
    \country{Austria}%
}
\email{otmar.ertl@dynatrace.com}

\author{Rick Rabiser}
\orcid{0000-0003-3862-1112}
\affiliation{%
    \institution{Johannes Kepler University Linz}%
    \city{Linz}%
    \country{Austria}%
}
\email{rick.rabiser@jku.at}

\begin{abstract}
    Performance benchmarking is a common practice in software engineering, particularly when building large-scale, distributed, and data-intensive systems. While cloud environments offer several advantages for running benchmarks, it is often reported that benchmark results can vary significantly between repetitions---making it difficult to draw reliable conclusions about real-world performance.
    
    In this paper, we empirically quantify the impact of cloud performance variability on benchmarking results, focusing on stream processing applications as a representative type of data-intensive, performance-critical system.
    In a longitudinal study spanning more than three months, we repeatedly executed an application benchmark used in research and development at Dynatrace.
    This allows us to assess various aspects of performance variability, particularly concerning temporal effects. With approximately 591~hours of experiments, deploying 789~Kubernetes clusters on AWS and executing 2\,366~benchmarks, this is likely the largest study of its kind and the only one addressing performance from an end-to-end, i.e., application benchmark perspective.

    Our study confirms that performance variability exists, but it is less pronounced than often assumed (coefficient of variation of~$<3.7\%$). Unlike related studies, we find that performance does exhibit a daily and weekly pattern, although with only small variability~($\le 2.5\%$).
    Re-using benchmarking infrastructure across multiple repetitions introduces only a slight reduction in result accuracy~($\le2.5$ percentage points). 
    These key observations hold consistently across different cloud regions and machine types with different processor architectures.
    We conclude that for engineers and researchers focused on detecting \emph{substantial} performance differences~(e.g., $>5\%$) in their application benchmarks, which is often the case in software engineering practice, performance variability and the precise timing of experiments are far less critical.

\end{abstract}



\maketitle

\section{Introduction}\label{sec:introduction}

Benchmarking and performance testing in general is an integral part of many software engineering and research activities~\cite{Kounev2020,Hasselbring2021}.
Today it is very common to run benchmarks in the cloud, for example, because it is more convenient for the engineers and researchers, because local machines for benchmarking are not available, or because production systems do also often run in the cloud, making the benchmarking results more representative.
However, there are many pitfalls reported when executing performance experiments in the cloud~\cite{Papadopoulos2021,Abedi2017,Netto2018,Uta2020}.
In particular, it is cautioned that performance results exhibit a high variability compared to execution environments that provide a higher level of control.
This leads to the general assumption that many repetitions are required or, otherwise, that benchmark results can be reported with only a low level of confidence.

In this study, we empirically assess the significance of performance variability in the context of an application benchmark for a distributed stream processing system.
Applications benchmarks test the performance of an entire application or system at its interfaces and are contrasted with microbenchmarks, which test individual functions or methods. Motivated by our R\&D efforts at Dynatrace on analyzing massive amounts of observability data in near-real time, we focus on the special case of distributed stream processing applications. Such systems are characterized by high performance requirements~\cite{SEAA2023}, making performance results crucial for development and operation decisions.

For this purpose, we combine the software engineering research methods \emph{benchmarking} and \emph{case study research}~\cite{Stol2018,Hasselbring2021}.
Through this research design, our study provides an in-depth analysis of cloud performance variability for a specific realistic application scenario. This contrasts with related work conducting broader, more generalized studies on the system or microbenchmark level~\cite{Leitner2016,Laaber2019,Baresi2023}.
The subject of our research is ShuffleBench~\cite{ICPE2024}, an open-source benchmark for distributed stream processing developed at Dynatrace. It is used in our research on assessing quality attributes such as performance, scalability, and resiliency of different implementation, configuration, and deployment alternatives.

In summary, we conducted 591~hours of experiments over a period of more than three months, in which we deployed 789~Kubernetes clusters in AWS and executed 2\,366~application benchmark runs. 
To the best of our knowledge, this is the largest study of this kind and the only one addressing performance from an end-to-end, i.e., application benchmark, perspective.
With these results, we are able to draw an accurate picture of cloud performance variability for the case of distributed stream processing applications.

\subsubsection*{Research questions}

Precisely, we address the following research questions with this study:

\begin{enumerate}[align=parleft,font=\bfseries]
	\item[RQ\,1] How high is performance variability for the case of our selected application benchmark?
	\item[RQ\,2] Does performance variability expose a daily pattern?
	\item[RQ\,3] Does performance variability expose a weekly pattern?
	\item[RQ\,4] Does performance variability expose a long-term pattern over multiple weeks?
	\item[RQ\,5] Is there a higher variability if the benchmarking infrastructure is re-provisioned and re-deployed between repetitions?
	\item[RQ\,6] Is performance variability impacted by the cloud machine type, in particular, with different processor architectures?
	\item[RQ\,7] Is performance variability impacted by the cloud region?
\end{enumerate}

\subsubsection*{Contributions}

In summary, we provide the following contributions:

\begin{itemize}
    \item A comprehensive study of performance variability over time for the case of stream processing application benchmarks, providing new insights by answering our stated research questions. It helps software engineering practitioners and researchers to decide when and how often to run their application benchmarks in the cloud.
    \item An experiment design serving as a template for other researchers and practitioners to repeat our research for the case of other benchmarks and other types of systems.
    \item An open dataset containing all our measurements as well as the complete source code of our data analysis~\cite{ReplicationPackage}. It allows other researchers to replicate and extend our research, for example, by conducting further analysis with our collected performance data.
\end{itemize}

\subsubsection*{Outline}

We start the remainder of this paper by discussing the background and related work on cloud performance variability in \cref{sec:background}.
\cref{sec:case-description} describes the subject of our study.
\cref{sec:experiment-design} describes the experiment design.
\cref{sec:experiment-results} presents and discusses our experimental results.
\cref{sec:threats-to-validity} discusses threats to validity and
\cref{sec:conclusions} concludes this paper.

\section{Background and Related Work}\label{sec:background}

Performance benchmarking is a common activity in the software development process, in particular, when building large-scale, distributed, and data-intensive software systems where high performance is essential. For example, at Dynatrace, we run various kinds of benchmarks to evaluate different architecture alternatives, implementation variants, and configuration options~
\cite{Ertl2024a,Reichinger2024,ICPE2024}.
Often benchmarks are classified as micro-benchmarks (similar to unit tests) and application benchmarks (similar to integration tests).

\subsubsection*{Cloud benchmarking}
With the advent of cloud computing, it has become quite common to also conduct performance experiments in the cloud~\cite{Schad2010,Folkerts2013,Bermbach2017}.
While software performance experiments in general already exhibit large variability in their results for various reasons, running experiments in public cloud environments reinforce this further~\cite{Abedi2017,Maricq2018}.
Typical reasons for this are effects of potential changes in the underlying hardware and the fact that software of different tenants runs on the
same hardware and, thus, interfere with each other. %
This is also referred to as the ``noisy neighbor'' problem~\cite{gkatzikis2013,Leitner2016}.
The suggested approach to cope with varying performance results is usually to repeat experiments and perform a statistical evaluation~\cite{Papadopoulos2021,Maricq2018,Bulej2017}, potentially with re-allocating cloud services %
and at different times.
However, it is often desirable to keep the overall execution time of benchmarks short to reduce costs and, increasingly, to lower carbon emissions as well.
A couple of methodologies~\cite{He2019,Bulej2020,He2021,Zhao2021} have been proposed by academia to achieve accurate performance results, while minimizing the number or execution time of experiments (e.g., by determining reliable conditions for terminating experiments early). Although promising, we find these methodologies challenging to apply in a real-world industry setting. Instead, our research is motivated by the need to assess how significant and practical relevant performance variability actually is when benchmarking applications in the cloud.

\subsubsection*{Cloud performance variability evaluation}

Over the past two decades, a couple of studies have been conducted to empirically assess and quantify the variability of performance in the cloud.
A frequently referenced work is that of \citet{Leitner2016}, in which the authors find that multi-tenancy has a major influence on performance variability. %
While the authors observe an influence of the cloud region on cloud performance, they find no clear impact of the time of day or the day of the week.
An earlier study by \citet{Iosup2011}, on the other hand, was able to recognize a yearly and daily pattern, although these results concern the APIs of cloud services.
\citet{Laaber2019} investigated how strong the performance variability of microbenchmarks in the cloud is and how reliable benchmark results can therefore be considered.
In a more recent study, \citet{Baresi2023} run various system-level benchmarks repeatedly over one month and use machine learning techniques to predict performance. They identified a  slight correlation between performance and both the time of day and weekends, but found no impact from the specific day of the week. Unlike our work, they do not quantify the impact of specific times of the day or the week.
Recently, performance variability has also been studied for Function-as-a-Service (FaaS) offerings. 
\citet{Schirmer2023} observed up to 15\% longer benchmark runtimes during working hours, whereas \citet{Eismann2022} found short-term and long-term performance changes when repeatedly executing a FaaS benchmark over 10~months.
\citet{Wen2025} discovered that the significant performance variability of FaaS benchmarks undermines the reliability of many existing benchmarking studies.
Although those works contribute tremendously to the understanding of cloud performance variability, our study differs from them in the following key aspects: 

    \emph{Age of studies.}
    Several papers that are frequently used as references for cloud performance variability date back to the earlier days of cloud computing \cite{Schad2010,Iosup2011,Leitner2016} and it is unclear to what extent their findings are still valid. Given the rapid evolution in the field of cloud computing, \citet{Leitner2016} called on the scientific community to periodically re-examine its understanding of the topic of cloud performance.

    \emph{Time span of investigation.}
    Compared to our work, fluctuations in cloud performance are often investigated over shorter periods of time~\cite{Schad2010,Leitner2016, Baresi2023}. This makes it more difficult to observe temporal patterns. \citet{Leitner2016} emphasizes the need for a longitudinal study that tracks performance over several months.

    \emph{Scope of benchmarks.}
    Most importantly, related studies mainly focus on micro-benchmarks or system-level benchmarks on Infra\-structure-as-a-Service virtual machines \cite{Laaber2019,Leitner2016,Baresi2023}. While such benchmarks are relevant to understand, for example, the behavior of the underlying execution infrastructure, they provide only limited insights into the performance of a specific application or software system. In contrast, we investigate performance variability for the case of an application benchmark. A key difference is that our application benchmark scenario is a distributed system consisting of different components, services, and middlewares communicating over the network. 
    As a result, our benchmark takes significantly longer to execute (i.e., several minutes compared to execution times of less than a second~\cite{Laaber2019,Eismann2022,Schirmer2023}).

\subsubsection*{Stream processing benchmarking}
Stream processing applications play a crucial role in enabling near-real-time data analytics across various domains such as finance, e-commerce, IoT, and software monitoring. There exists a variety of frameworks, such as Apache Kafka Streams~\cite{Wang2021}, that enable building such applications as distributed systems that process large volumes of data with low latency.
Performance of those applications is often critical~\cite{SEAA2023} and, therefore, many stream processing benchmarks have been proposed and performance evaluation studies have been conducted by both academia and industry~\cite{SEAA2023,Hesse2021,JSS2024,ICPE2024}.
Throughput is particularly important as a higher throughput per instance reduces the number of instances required to handle a given workload, thereby lowering the associated costs of running them in the cloud.
Performance variability of stream processing applications has not been studied much, except for performance variations due to varying workloads. However, related to this study is the work of \citet{EMSE2022} that evaluates how often and for how long a stream processing benchmark should be executed in the cloud to provide sufficient confidence on whether a certain load intensity is processable or not.
Instead of this binary decision, however, our study investigates how high the variability of the measured throughput is, in particular with regard to the execution time of the experiment as well as potential influences by the virtual machine types and cloud region.

\section{Case Description}\label{sec:case-description}

To address our stated research questions, we combine the empirical software engineering research methods \emph{benchmarking} and \emph{case study}.\footnote{We are aware of the controversy around the term ``case study research''~\cite{Wohlin2022}. Our study shares many properties typically required by case study guidelines such as investigating a contemporary phenomenon in depth and within its real-world context. However, as such guidelines mainly address the investigation of social phenomena with quantitative analysis, we would not classify our study as a case study per se.}
Benchmarking is both a software engineering research method and a common activity in software engineering practice.
We use benchmarking to obtain performance measurements in a representative and reproducible way. However, in contrast to standard benchmarking studies, our goal is not to compare different alternatives with each other. Instead, we conduct a kind of ``meta study'', investigating the variability of application benchmark runs in the cloud.
For this purpose, we adopt an approach inspired by a longitudinal, evaluative case study.\footnote{\url{https://www2.sigsoft.org/EmpiricalStandards/docs/standards?standard=CaseStudy}} Specifically, we conduct an in-depth study of the phenomenon of cloud performance variability in a real-world context by focusing on a specific application scenario. %
In this section, we describe the subject of our study in detail.

\subsection{Distributed Stream Processing Applications}
With our study we focus on stream processing applications, which process continuous streams of data with low (often sub-second) latency. By filtering, transforming, or aggregating records in data streams, such applications facilitate near-real-time data analytics across diverse domains, including finance, e-commerce, IoT, and software monitoring.
Due to ever-growing volumes of data and the widespread availability of scalable compute resources through cloud platforms, stream processing applications are often implemented as distributed systems. 
This introduces the need to address critical properties such as fault-tolerance, scalability, resource efficiency, state management, and data partitioning.
State-of-the-art open-source stream processing frameworks and systems provide advanced mechanisms to support these properties, enabling organizations to build high-performance distributed applications. However, numerous challenges can arise during design, implementation, or deployment, making regular performance evaluation essential.

\subsection{Kubernetes-based Execution Environment}
Cloud providers offer a wide range of services, providing different abstraction levels. In our study, we focus on a managed Kubernetes environment consisting of virtual machine nodes. This level of abstraction is a common choice for operating large-scale, distributed, and data-intensive software systems in the cloud, as done, for example, for many parts of the Dynatrace platform. For our study, we select the largest cloud provider Amazon Web Services (AWS) with its Elastic Kubernetes Service (EKS) offering.

\subsection{The ShuffleBench Application Benchmark}

As subject of our evaluation, we use ShuffleBench~\cite{ICPE2024}, our open-source\footnote{\url{https://github.com/dynatrace-research/ShuffleBench}} benchmark for distributed stream processing frameworks.
ShuffleBench is inspired by requirements for near real-time analytics of observability data at Dynatrace. Yet, by focusing on the core use case of \textit{shuffling} (i.e., re-distributing) data records to perform state-local aggregations, we expect performance results obtained by ShuffleBench to be representative of many real-world applications.
At Dynatrace, ShuffleBench is used to evaluate different design decisions for processing streams of observability data at large scale~\cite{ICPE2024,DEBS2024,SANER2025}.

In contrast to several other stream processing benchmarks, ShuffleBench is an application benchmark that involves heavy usage of CPU, memory, network, and disk as well as access to managed cloud services. It provides ready-to-use implementations for a set of state-of-the-art stream processing frameworks and leverages Apache Kafka as source and sink for data streams, a messaging system widely adopted in industry.
Additionally, ShuffleBench comes with a load generator and deployment definitions in the form of Kubernetes manifests.
ShuffleBench is provided as an executable benchmark for the cloud-native benchmarking framework Theodolite\footnote{\url{https://www.theodolite.rocks/}}~\cite{EMSE2022}. This automates the benchmarking process, including deployment and deletion of all benchmark components in Kubernetes, monitoring, and data collection.

While ShuffleBench also supports measuring qualities such as latency, scalability, and fault recovery time, we focus on measuring throughput according to the ad-hoc measurement method~\cite{ICPE2024}. The obtained throughput values provide a good estimate of the load a similar real-world application could sustain under typical operating conditions.
For us at Dynatrace, achieving a high throughput is particularly relevant to cope with immense load on the platform, while minimizing the required computing resources for processing.

\subsection{The Kafka Streams Framework}

Apache Kafka Streams \cite{Sax2018,Wang2021} is a popular framework for implementing distributed stream processing applications. It is tightly integrated with the Apache Kafka messaging system and can be embedded as a library in a Java-based microservice, providing automatic coordination among service instances, scalable data partitioning, state management, and fault tolerance.
Although ShuffleBench supports different state-of-the-art stream processing frameworks, we choose the Kafka Streams implementation for our study due to its industry popularity and relevance for Dynatrace for cloud-native, microservice-like applications.

\section{Experiment Design}\label{sec:experiment-design}

The core idea of our experiment design is to periodically execute the same benchmark.
This allows us to assess variability across several runs and rules out external influences to a great extent. To investigate whether variability exposes temporal patterns, we run the benchmark every day at the same times of the day, allowing to group and aggregate benchmark results by the same hour of day, the same day of the week, or the same week.

\subsection{Automated Benchmarking Process}
We designed an automated process in AWS, which periodically sets up the benchmark environment, runs the benchmark, and collects the benchmark results data.
Starting point of this periodic process is a scheduled task in AWS Elastic Container Service (ECS). Whenever executed, this task creates a new EKS Kubernetes cluster and installs the benchmarking infrastructure in this cluster, including Apache Kafka, monitoring tooling, and the Theodolite benchmarking framework.
Once everything is set up, the ECS task initiates the execution of the benchmark through Theodolite with a configurable number of repetitions. Afterwards, all benchmark results are copied to an AWS Simple Storage Service (S3) bucket for later analysis, before the benchmarking infrastructure is uninstalled and the cluster is deleted again.

\subsection{Benchmark Configuration}\label{sec:experiment-design:configuration}

In general, we apply the same benchmark configuration and execution infrastructure as in our previous study~\cite{ICPE2024}.
That means, for example, that our Kubernetes cluster is provisioned in the AWS \emph{us-east-1} region and consists of 10 nodes, (i.e., EC2 virtual machine instances). The cluster is divided into 3~\emph{m6i.xlarge} nodes hosting the stream processing framework, 3~\emph{m6i.2xlarge} nodes hosting one Kafka broker each, and 4~\emph{m6i.xlarge} nodes that host the load generator instances plus additional benchmarking infrastructure. In addition to these \emph{baseline} experiments, we conduct a set of experiments with \emph{m6g} instances to address \RQ{6} and a set of experiments in the \emph{eu-central-1} region to address \RQ{7}.
Note that in contrast to the ShuffleBench results of our previous publication~\cite{ICPE2024}, we run these experiments in a single availability zone to reduce the costs for network traffic. This leads to slightly better performance results compared to our previous study.

The ShuffleBench Kafka Streams implementation is deployed with 9 application instances (3 per cluster node), where each instance is assigned 4\,GB of memory and one virtual CPU core, resulting in a total parallelism of 9.

We measure the achievable throughput of a stream processing application according to the ad-hoc throughput method~\cite{ICPE2024}: We generate a constant high load on the system (1~million data records per second in this case) and continuously monitor how many of those records could be processed per second. Everything that is not processed queues up in the messaging system Apache Kafka. We execute those experiments over a period of 15~minutes, which we found to be sufficient to capture the fluctuating nature of throughput~\cite{ICPE2024}.
With throughput measurements taken every 5~seconds, this results in 180~data points per experiment. We consider the first 3~minutes as warm-up period, in which we found the throughput to be less stable~\cite{ICPE2024}, and removed to corresponding data points. Afterward, we average throughput over the remaining duration to smooth out fluctuations and provides a representative measure of overall system performance.
For the subsequent analysis in \cref{sec:experiment-results}, we consider these average throughput values as the results of the individual benchmark executions.

\subsection{Periodic Benchmark Execution}

We configured the periodic benchmarking task to be executed every 6~hours to cover a full day cycle. For a period of three weeks, we additionally decreased the time between experiments to 3~hours to get an indication of a more fine-grained daily pattern (see \cref{sec:experiment-results:daily}).
Within one periodic task execution, the benchmark is executed 3~times to incorporate performance variability within the same infrastructure (see research question \RQ{5}).
An overview of all benchmark executions is shown in \cref{tab:timeframes}.
Provisioning and deleting cloud resources on AWS, in particular EC2 instances, take a considerable amount of time. Hence, a periodic task runs for roughly 1:20 hour, which is considerably longer than the pure benchmark execution time of only 0:45 hours.

\begin{table}
	\caption{Time periods of experiments conducted in this study.}
	\label{tab:timeframes}%
	\renewcommand{\arraystretch}{1.2}
	\begin{tabularx}{\linewidth}{@{}lXlrr@{}}
		\toprule
		AWS region & EC2 & Time period & Days & Execs. \\
		\midrule
		us-east-1 & m6i & 2024-05-14 -- 2024-07-29 & 76 & 1086 \\
		 &  & 2024-09-24 -- 2024-10-01 & 7 & 81 \\
		us-east-1 & m6g & 2024-06-04 -- 2024-07-29 & 55 & 630 \\
		eu-central-1 & m6i & 2024-09-16 -- 2024-11-03 & 48 & 569 \\
		\midrule
		 &  & \multicolumn{1}{r}{Total:} & 124 & 2366 \\
		\bottomrule
	\end{tabularx}
\end{table}

\section{Experiment Results}\label{sec:experiment-results}

In the following, we present and discuss the results of our experiment with respect to our research questions \RQ{1--7}.

\subsection{General Performance Variability}

While we can assume from the existing literature that there is significant variability in performance in the cloud, our research question \RQ{1} aims to quantify this variability for our studied application type, which we expect to be more representative for real-world workloads compared to related studies.
To address this research question we summarize all results of our baseline configuration.

\begin{figure}%
	\includegraphics[width=\linewidth]{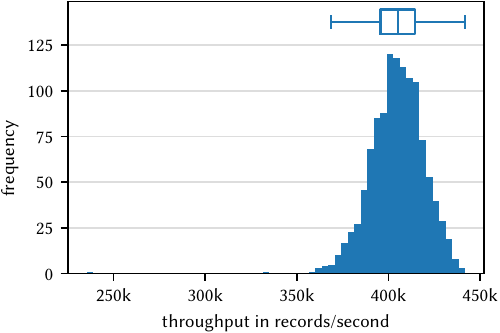}%
	\caption{Histogram and associated boxplot of all throughput measurements in our baseline experiments.}
	\label{fig:histogram}
\end{figure}

\Cref{fig:histogram} shows a histogram with an associated boxplot of the observed throughput results. As common for performance data, the distribution is slightly left-skewed~\cite{Kounev2020}.
In particular, we notice two clear outliers with significantly lower throughput.
The associated summary statistics of this baseline evaluation are listed in \cref{tab:summary-statistics}.
The histogram already indicates that the data is not normal distributed, which we confirm with a Shapiro--Wilk test ($p < 0.001$). However, as the distribution shows a clear central tendency, almost symmetry in the interquartile range, and very similar mean and median values, we focus for the following analyses on statistics based on the arithmetic mean and standard deviation.

The coefficient of variation (CV) is a measure to relate the mean and the standard deviation. Also referred to as relative standard deviation, it is used in the related literature to quantify cloud performance variability~\cite{Schad2010,Leitner2016,Laaber2019,Uta2020,Eismann2022,Wen2025}.
In our baseline experiments, the CV is 3.69\%, which is on the lower end of the wide range of variability reported for micro and system-level benchmarks in the literature.
More intuitively, we find that 50\% of all measurements are within $-2.4\%$ and $+2.3\%$ of the median (i.e., the interquartile range). %
However, caution should be paid to the outliers: Although very rare (less than $0.2\%$ of all executions), we could observe benchmark executions in which the result deviates extremely from the expected value.

\begin{resultsbox}{\RQ{1}}
    Cloud performance variability clearly exists, but contrary to what is sometimes assumed, it is not inherently detrimental when benchmarking on the application level.
    In cases where a few percentage points of uncertainty can be tolerated, extensive repetitions---as common in micro-benchmarking---are not required for the case of our application benchmark.
\end{resultsbox}

\subsection{Daily Pattern}\label{sec:experiment-results:daily}

\begin{figure}%
	\begin{subfigure}[c]{\linewidth}%
		\includegraphics[width=\linewidth]{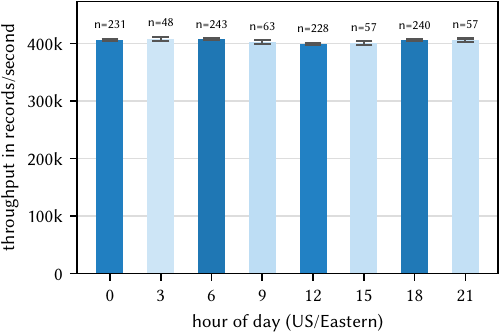}%
		\caption{Full y-axis range}%
	\end{subfigure}
        
        \figvspace
        
	\begin{subfigure}[c]{\linewidth}%
		\includegraphics[width=\linewidth]{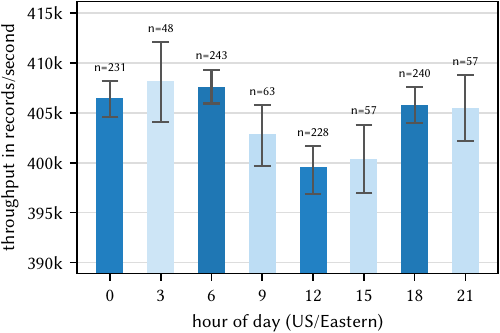}%
		\caption{Zoomed-in y-axis range}%
	\end{subfigure}
	\caption{Measured throughput summarized by the hour of the day. The color intensity indicates the sample size.}
	\label{fig:daily}
\end{figure}

With our research question \RQ{2}, we are interested in whether the expected result of a benchmark execution depends on the time of the day the benchmark is executed. In particular, we investigate whether the performance variability exposes a daily pattern.
To address \RQ{2}, we summarize all results by the hour of the day when the corresponding experiment was executed.
The mean observed throughput per hour of day with its corresponding confidence intervals (obtained via bootstrapping) are shown in \cref{fig:daily}.
As mentioned in \cref{sec:experiment-design:configuration},
we conducted experiments every three hours instead of every six hours over a few weeks to get a clearer picture of the daily course. All our results are shown in US/Eastern timezone, which corresponds to the physical location of AWS' region \emph{us-east-1}.
Our results show that the mean throughput remains very similar over the course of the day, however, zooming in we can see a clear daily pattern. The lowest throughput can be observed at noon (around 12:00), while the highest throughput is achieved at around 3:00 to 6:00 with a difference of the mean of 2.15\%.

However, the large overlap in confidence intervals in \cref{fig:daily} suggests that these results should be interpreted with caution. %
We additionally performed pair-wise Mann--Whitney U tests to see whether the throughput results at certain hours of the day are significantly different from another. \Cref{tab:daily} shows the test's p-values. Using a 95\% confidence level, we consider differences between different hours of the day statistically significant if $p \le 0.05$ (highlighted in bold).
Hence, we consider the results from 0:00 to 6:00 significantly different from the results from 9:00 to 15:00 and the results from 12:00 to 15:00 significantly different from the results from 18:00 to 21:00.

\begin{table}
	\caption{Mann--Whitney U test's p-values to assess whether there is a statistically significant difference between the results at different hours of the day.}
	\label{tab:daily}
	\small
	\begin{tabularx}{\linewidth}{@{}lRRRRRRRR@{}}
		\toprule
		 & 0 & 3 & 6 & 9 & 12 & 15 & 18 & 21 \\
		\midrule
		0 &  & .408 & .591 & \textbf{.035} & \textbf{\textless .001} & \textbf{.001} & .489 & .580 \\
		3 & .408 &  & .621 & \textbf{.025} & \textbf{\textless .001} & \textbf{.003} & .243 & .228 \\
		6 & .591 & .621 &  & \textbf{.021} & \textbf{\textless .001} & \textbf{\textless .001} & .234 & .360 \\
		9 & \textbf{.035} & \textbf{.025} & \textbf{.021} &  & .112 & .259 & .142 & .207 \\
		12 & \textbf{\textless .001} & \textbf{\textless .001} & \textbf{\textless .001} & .112 &  & .936 & \textbf{\textless .001} & \textbf{.005} \\
		15 & \textbf{.001} & \textbf{.003} & \textbf{\textless .001} & .259 & .936 &  & \textbf{.007} & \textbf{.023} \\
		18 & .489 & .243 & .234 & .142 & \textbf{\textless .001} & \textbf{.007} &  & .949 \\
		21 & .580 & .228 & .360 & .207 & \textbf{.005} & \textbf{.023} & .949 &  \\
		\bottomrule
	\end{tabularx}
\end{table}

\begin{resultsbox}{\RQ{2}}
    Our analysis reveals a subtle yet statistically significant daily pattern in performance. Benchmarks executed around noon tend to exhibit slightly lower performance, whereas those conducted during late-night and early-morning hours achieve the highest results.
\end{resultsbox}

\subsection{Weekly Pattern}\label{sec:experiment-results:weekly}

\begin{figure}%
	\begin{subfigure}[c]{\linewidth}%
		\includegraphics[width=\linewidth]{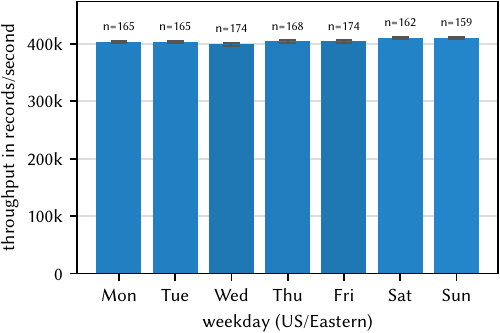}%
		\caption{Full y-axis range}%
	\end{subfigure}
        
        \figvspace
        
	\begin{subfigure}[c]{\linewidth}%
		\includegraphics[width=\linewidth]{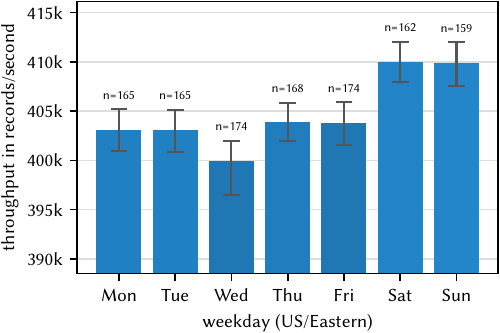}%
		\caption{Zoomed-in y-axis range}%
	\end{subfigure}
	\caption{Measured throughput summarized by the day of the week. The color intensity indicates the sample size.}
	\label{fig:weekly}
\end{figure}

Similar to the previous analysis, research question \RQ{3} asks if the weekday has impact on the performance of benchmark runs.
We follow a similar approach as for the daily pattern: We summarize all our baseline experiment results by the day of the week they have been executed. The corresponding mean throughput result with the associated bootstrapped confidence intervals are depicted in \cref{fig:weekly}. Although the absolute values do not show clear differences, a closer look reveals that performance on weekend runs is considerably higher than on weekdays. Additionally, the performance on Wednesdays is slightly lower than on the other days. The maximum variability is similar to the daily pattern with a difference of 2.52\% in mean throughput from Saturdays to Wednesday.

To get a second indicator of whether the observed differences are statistically significant, we again performed pair-wise Mann--Whitney U tests. \Cref{tab:weekly} shows the test's p-values. Using a 95\% confidence level, we conclude that the results at Saturdays and Sundays differ significantly from the results at weekdays and, additionally, the results at Wednesdays differ significantly from those of Thursdays and Fridays.

\begin{table}
	\caption{Mann--Whitney U test's p-values to assess whether there is a statistically significant difference between the results at different weekdays.}
	\label{tab:weekly}
	\small
	\begin{tabularx}{\linewidth}{@{}lRRRRRRR@{}}
		\toprule
		& Mon & Tue & Wed & Thu & Fri & Sat & Sun \\
		\midrule
		Mon &  & .846 & .085 & .725 & .679 & \textbf{\textless .001} & \textbf{\textless .001} \\
		Tue & .846 &  & .054 & .850 & .793 & \textbf{\textless .001} & \textbf{\textless .001} \\
		Wed & .085 & .054 &  & \textbf{.024} & \textbf{.033} & \textbf{\textless .001} & \textbf{\textless .001} \\
		Thu & .725 & .850 & \textbf{.024} &  & .991 & \textbf{\textless .001} & \textbf{\textless .001} \\
		Fri & .679 & .793 & \textbf{.033} & .991 &  & \textbf{\textless .001} & \textbf{\textless .001} \\
		Sat & \textbf{\textless .001} & \textbf{\textless .001} & \textbf{\textless .001} & \textbf{\textless .001} & \textbf{\textless .001} &  & .618 \\
		Sun & \textbf{\textless .001} & \textbf{\textless .001} & \textbf{\textless .001} & \textbf{\textless .001} & \textbf{\textless .001} & .618 &  \\
		\bottomrule
	\end{tabularx}
\end{table}

\begin{resultsbox}{\RQ{3}}
    We observe a modest weekly pattern in performance. Benchmarks executed over the weekend show slightly higher performance compared to weekdays, with Wednesday standing out as the day with the lowest performance.
\end{resultsbox}

\subsection{Long-Term Pattern}\label{sec:experiment-results:long-term}

For \RQ{4}, we are interested in whether the performance shows long-term patterns or trends. To answer it, we performed a similar analysis as done before, but summarizing all measurements by the ISO week number of the experiment dates. In total, 11~weeks of measurements are included in this analysis.

\begin{figure}%
	\includegraphics[width=\linewidth]{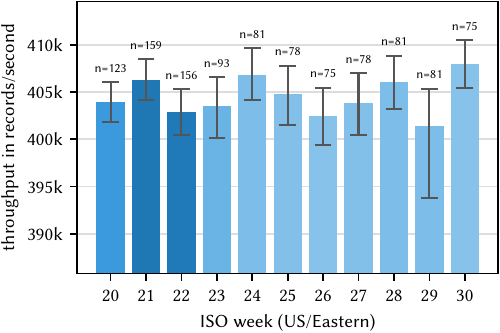}%
	\caption{Measured throughput summarized by the ISO week number with zoomed-in y-axis range. The color intensity indicates the sample size.}
	\label{fig:yearly}
\end{figure}

\Cref{fig:yearly} shows the mean throughput for each week with its corresponding bootstrapped confidence interval. Checking for overlapping confidence intervals as well as pair-wise Mann--Whitney U tests (with 95\% confidence level) indicate that only the last week shows a statistically significant difference to a few other weeks.
Hence, from our data we cannot conclude that there are long-term pattern throughout the year. However, our experiments span only a part of the year, so we cannot rule out that there is a difference during other periods.

\begin{resultsbox}{\RQ{4}}
    We observe small performance fluctuations over time, yet our results provide no indication of a long-term pattern or trend.
\end{resultsbox}

\subsection{Impact of Re-using Infrastructure}

Setting up cloud infrastructure can take a considerable amount of time and, hence, cause additional costs. %
For example, provisioning the Kubernetes cluster for our benchmark takes around 20~minutes. Research question \RQ{5} concerns whether there is a significant difference in the performance results if we re-provision a cluster for different repetitions opposed to re-using the same infrastructure across multiple repetitions.

There are multiple levels at which benchmark infrastructure can be re-used. For our experiments, we always conduct three benchmark runs in the same EKS cluster, with the same underlying virtual machines, and the same benchmarking infrastructure installation including the Kafka cluster, monitoring tooling, and benchmark orchestration.

\begin{figure}%
	\includegraphics[width=\linewidth]{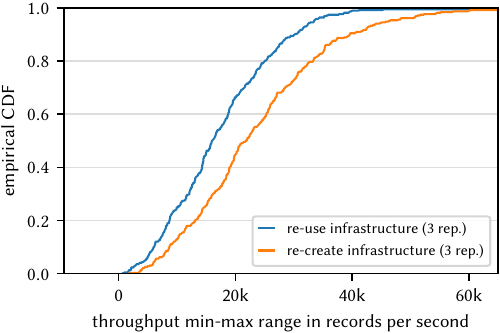}%
	\caption{ECDF of difference between minimum and maximum of three observed value when re-using the benchmark infrastructure and when re-creating it before each execution.}
	\label{fig:ptp}
\end{figure}

We contrast the variability across these three benchmark runs using the same infrastructure with the variability of randomly sampled three benchmark runs, which represents benchmark runs with re-created infrastructure.
To quantify variability, we use the min-max range---that is, the difference between the maximum and minimum throughput across the three repetitions.
\Cref{fig:ptp} shows an empirical cumulative distribution function (eCDF) of the min-max ranges of all sets of executed benchmarks that re-use the same infrastructure in comparison to an eCDF of the min-max ranges of the randomly sampled benchmark runs (i.e., with re-created infrastructure).
We can see that re-creating the benchmark infrastructure leads to higher variability. A Kolmogorov--Smirnov test confirms this observation ($p < 0.001$), indicating that the benchmark infrastructure has a significant impact on the results.

\begin{figure}%
	\includegraphics[width=\linewidth]{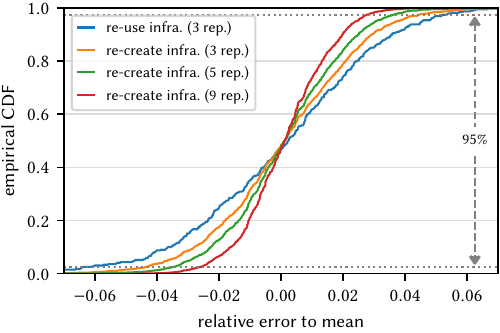}%
	\caption{ECDF of relative errors when re-using the benchmark infrastructure and when re-creating it before each execution with numbers of repetitions.} %
	\label{fig:error}
\end{figure}

Moreover, we quantify the expected error induced by re-using the same benchmarking infrastructure across repetitions and compare it to the expected error when re-creating the infrastructure.
For this purpose, we compute for each set of repetitions the relative error of the median throughput of the three repetitions to the true mean (i.e., the mean across all experiments). %
In addition, we also compute the relative errors for re-creating the infrastructure with 5 and 9 repetitions.
\Cref{fig:error} shows an eCDF of the individual errors.
One can note that if the benchmark infrastructure is re-used and experiments are repeated three times, the median throughput of these repetitions is within [-6.4\%, 5.2\%] of the true mean in 95\% of all cases.
In the case of re-creating the benchmark infrastructure between each repetition, a lower error of only [-4.6\%, 4.5\%] is achieved.
Naturally, repeating experiments more often can reduce the error.
For example, with 5~repetitions we obtain an expected error of [-3.5\%, 3.62\%], while 9~repetitions further decrease it to [-2.62\%, 2.79\%].

\begin{resultsbox}{\RQ{5}}
    Re-using the same benchmarking infrastructure across different repetitions of a benchmark execution is appealing because it reduces the overall benchmarking time. However, one should be aware of the associated increased error range, which is in our case study $2.5$ percentage points larger when compared to re-provisioning and re-deploying the benchmarking infrastructure for each repetition.
\end{resultsbox}

\subsection{Impact of Cloud Machine Type}

We conduct our baseline experiments with \emph{m6i} instances, which are considered a good choice for general purpose workloads with moderate costs. It is reasonable to assume that this instance type family is among the most used.
For research question \RQ{6}, we investigate whether the previously observed performance variability also occurs with other machine types. For this purpose, we repeat our experiments with \emph{m6g} instances. While \emph{m6i} instances are powered by Intel Xeon x86 processors, \emph{m6g} instances use ARM-based Graviton2 processors designed by AWS.
ARM-based cloud machines have gained significant attention recently due to their advertised cost-effectiveness and energy efficiency.
Contrasting our previous performance variability results with \emph{m6g} instances is interesting as they are likely less utilized and their distinct processor architecture could introduce unique performance dynamics.

\begin{figure}%
	\begin{subfigure}[c]{0.485\linewidth}%
		\includegraphics{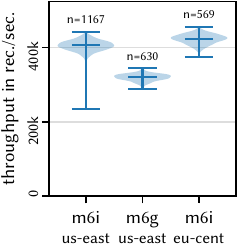}%
		\caption{Throughput}%
		\label{fig:comparison:throughput}
	\end{subfigure}%
	\hfill%
	\begin{subfigure}[c]{0.485\linewidth}%
		\includegraphics{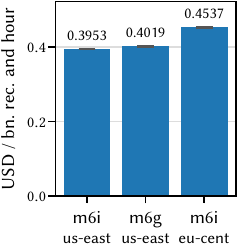}%
		\caption{Costs per billion records}%
		\label{fig:comparison:costs}
	\end{subfigure}
	\caption{Comparison of throughput and costs across two different instance types and and two different cloud regions.}
	\label{fig:comparison}
\end{figure}

\Cref{fig:comparison:throughput} shows a comparison of the frequency distributions of our experiments with the \emph{m6g} and the \emph{m6i} instance. Corresponding statistical attributes are summarized in \cref{tab:summary-statistics}. The measured throughput with \emph{m6g} instances is significantly slower, but follows a similar distribution. We observe a slightly lower variability with a CV of 2.92\%.
While the benchmark executions on \emph{m6g} instances yield significantly lower throughput (by $21.1\%$), also the costs charged by AWS are lower for \emph{m6g} instances than for \emph{m6i} instances. \Cref{fig:comparison:costs} compares the costs per hour (in USD following AWS list prices) for processing one billion records. It shows that the costs for processing the same volume of data is almost identical with the \emph{m6i} instances being 1.6\% cheaper.

\begin{table*}
    \caption{Summary statistics of measured throughput in records/second across two different instance types and two different cloud regions (IQR: interquartile range, Q1: first quartile, Q3: third quartile, CV: coefficient of variation).}
    \label{tab:summary-statistics}%
    \renewcommand{\arraystretch}{1.2}
    \begin{tabularx}{\linewidth}{@{}l X rrrrrrrrr@{}}
        \toprule
        AWS region &
        EC2 &
        Mean &
        Median &
        Min &
        Max &
        Std. Dev. &
        IQR	&
        Q1 &
        Q3 &
        CV \\
        \midrule
        us-east-1 &
        m6i &
        404\,743.7 &
        405\,376.0 &
        235\,405.3 &
        441\,835.1 &
        14\,922.0 &
        18\,941.6 &
        395\,638.0 &
        414\,579.6 &
        3.68\% \\
        
        us-east-1 &
        m6g &
        319\,332.1 &
        319\,924.3 &
        288\,260.8 &
        343\,849.3 &
        9\,347.4 &
        12\,612.6 &
        313\,664.4 &
        326\,277.0 &
        2.92\% \\
        
        eu-central-1 &
        m6i &
        422\,408.7 &
        423\,588.3 &
        374\,489.6 &
        455\,366.5 &
        13\,176.6 &
        17\,697.8 &
        413\,967.0 &
        431\,664.7 &
        3.11\%\\
        \bottomrule
    \end{tabularx}
\end{table*}

\begin{figure}%
	\begin{subfigure}[c]{\linewidth}%
		\includegraphics[width=\linewidth]{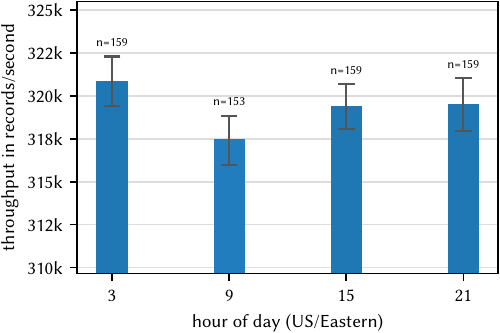}%
		\caption{Summarized by the hour of day}%
	\end{subfigure}

        \figvspace
    
	\begin{subfigure}[c]{\linewidth}%
		\includegraphics[width=\linewidth]{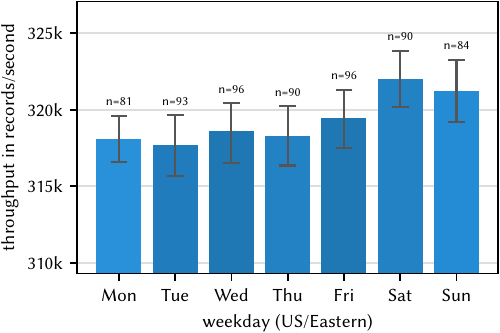}%
		\caption{Summarized by the day of the week}%
	\end{subfigure}
	\caption{Measured throughput with \emph{m6g} instances summarized by time attributes.}
	\label{fig:useast1-m6g}
\end{figure}

We also investigate whether running our application benchmark on \emph{m6g} instances reveals the same daily and weekly pattern we have seen for \emph{m6i} instances.
\Cref{fig:useast1-m6g} shows our benchmark executions summarized by the hour of the day and the day of week along with bootstrapped confidence intervals.
It indicates temporal patterns similar to those seen in~\cref{sec:experiment-results:daily} and~\cref{sec:experiment-results:weekly}, although a bit less pronounced and with some small differences.
However, the larger confidence intervals---likely due to the smaller sample size---reduce confidence in these results.

\begin{resultsbox}{\RQ{6}}
    We found no substantial difference in cloud performance variability between the widely used x86-based virtual machines and the emerging ARM-based instances. While the ARM-based instances deliver significantly lower throughput for our tested stream processing application, this is offset by their lower hourly cost, resulting in nearly equivalent costs per processed data volume.
\end{resultsbox}

\subsection{Impact of Cloud Region}

The \emph{us-east-1} region used in our previous results is likely the most widely used AWS region, as it was the first to launch, offers the largest number of availability zones, and provides the most significant set of services and features. As a result, it is a strong choice not only for businesses based in the eastern United States but also for global organizations that do not have stringent latency or compliance requirements.
With our research questions \RQ{7}, we aim to investigate whether our previous findings depend on the cloud region or can be replicated in another region. For this purpose, we selected the \emph{eu-central-1} region, which we expect to be commonly used by organizations based in central Europe.

\Cref{fig:comparison:throughput} and \cref{tab:summary-statistics} show that our benchmark's throughput is generally higher in the \emph{eu-central-1} region compared to \emph{us-east-1}, although both follow a similar distribution with a slightly lower variability according to a CV of 3.11\%.
The experiments in \emph{eu-central-1} were executed a few months after the initial experiments in \emph{us-east-1}. To check whether the higher performance could be due to a long-term change beyond our results in \cref{sec:experiment-results:long-term}, we repeated the experiments in \emph{us-east-1} over one week at the same time as the \emph{eu-central-1} experiments. We pairwise compared the results for both regions from the same time. As we found that in about 87\% of the cases, the \emph{eu-central-1} regions shows higher performance, we conclude that the difference is region-based rather than time-related.
It is worth noting that although \emph{eu-central-1} achieves higher performance, the processing costs remain 12.9\% lower in \emph{us-east-1} due to its lower per-instance pricing (see \cref{fig:comparison:costs}).
These findings also underscore the importance of reporting the cloud region used, which is not always done in benchmarking studies~\cite{Baresi2023,Eismann2022,EMSE2022}.

\begin{figure}%
	\begin{subfigure}[c]{\linewidth}%
		\includegraphics[width=\linewidth]{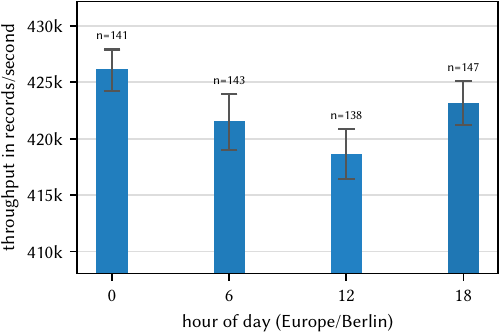}%
		\caption{Summarized by the hour of day}%
	\end{subfigure}

        \figvspace
    
	\begin{subfigure}[c]{\linewidth}%
		\includegraphics[width=\linewidth]{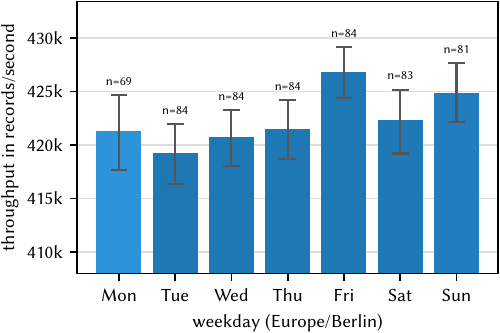}%
		\caption{Summarized by the day of the week}%
	\end{subfigure}
	\caption{Measured throughput in the \emph{eu-central-1} region summarized by time attributes.}
	\label{fig:eucentral1-m6i}
\end{figure}

\Cref{fig:eucentral1-m6i} shows the results from replicating the analyses from \cref{sec:experiment-results:daily} and \cref{sec:experiment-results:weekly} for the \emph{eu-central-1} region. Here, all results are reported in Europe/Berlin timezone, which corresponds to the physical location of the \emph{eu-central-1} region.
We observe a daily pattern similar to that in \emph{us-east-1}, with benchmark executions during the day showing lower performance than those at night.
Compared to the \emph{us-east-1} region, however, the highest performance appears in earlier hours of the night at around 0:00.
An interesting observation is that in \emph{eu-central-1}, Friday is the day with the highest performance on average, whereas in \emph{us-east-1} this is Saturday and Sunday. %
The performance drop on Saturdays is particularly remarkable and could not be observed in \emph{us-east-1}.

\begin{resultsbox}{\RQ{7}}
    The performance of the same machine type can vary across regions, meaning that benchmark results from different regions should only be compared with caution and, in particular, in relation to costs.
    While performance variability shows similar daily and weekly patterns in the \emph{eu-central-1} region, we observe small but notable differences compared to the \emph{us-east-1} region.
\end{resultsbox}

\section{Threats to Validity}\label{sec:threats-to-validity}

Despite careful research design, there are threats and limitations to the validity of our study, which we discuss below.

\subsubsection*{Internal Validity}

The primary objective of our study is to explore the temporal aspect of cloud performance variability. To achieve this, we focus on maximizing the number of benchmark executions across different times. This ensures robust data for temporal analysis, but limits the exploration of other potential variations in our experimental design and, hence, poses a threat to validity.
For example, we focused on AWS and can therefore not rule out that other cloud providers exhibit other performance variability patterns.
Related work~\cite{Leitner2016, Laaber2019, EMSE2022,Uta2020} does not provide sufficient clarity to make definitive assumptions about whether performance variability differs across cloud environments.
Likewise, we focused on Kafka Streams as stream processing framework. The performance of modern stream processing frameworks can differ significantly depending on the use case~\cite{SEAA2023}, although we observed no significant differences in variability in our previous work~\cite{ICPE2024}.
We increase the internal validity by repeating our experiments with a second type of virtual machines for the Kubernetes nodes as well as in a second cloud region. However, we cannot rule that fundamentally different machine types or regions behave differently. In particular, it is important to note that during the time of this study, AWS released a new generation of EC2 instance types, and we cannot be certain that our results apply to these newer instances. 

\subsubsection*{External Validity}

An inherent implication of our case study-like research design is its limited generalizability. With stream processing applications, our study subject is a representative type of data-intensive, performance-critical systems. However, other types of applications might be subject to other usage patterns or performance requirements, leading to distinct interactions with with the underlying cloud environment.
As previous research~\cite{Leitner2016,Laaber2019} found significant differences in performance variability when benchmarking on lower system levels, we assume this may also be reflected in some application benchmarks.
Nonetheless, the insights gained from our analysis of stream processing applications can serve as a valuable reference for understanding performance variability in other data-intensive and performance-critical systems, particularly those with similar cloud interactions and workload characteristics.

\section{Conclusions}\label{sec:conclusions}

This paper reports on our long-term study of empirically quantifying performance variability when running application benchmarks in the cloud.
Following a case study-like approach, we focus on stream processing applications with an open-source benchmark designed and frequently used at Dynatrace, representing a typical data-intensive, performance-critical system.

Our results confirm that application-level benchmark performance exhibits noticeable variability when run in the cloud.
However, the variability is less pronounced than often implied with a CV of less than 3.7\%. The distribution of performance measurements shows a clear central tendency with only slight skewness and very rare (less than 0.2\%) extreme outliers.
We observe subtle daily and weekly performance patterns: Benchmarks executed during the night exhibit slightly higher performance than those executed during the day, with differences of up to 2.1\%. Benchmarks executed on the weekend show slightly higher performance over weekday runs, with differences of up to 2.5\%.
From our experiments, we have no indication for a long-term, seasonal pattern.
Re-provisioning and re-deploying the benchmark infrastructure between benchmark repetitions can help to obtain slightly more accurate results (e.g., a 2.5 percentage points smaller error range with 3~repetitions), but come at the cost of longer experiment durations and higher costs.
Despite subtle differences, our key observations hold independently of the cloud region and the virtual machine type.

In response to the title of our study---``When should I run my application benchmark?''---the answer depends on the benchmarking objective:
If the goal is to detect small performance differences in the order of less than 5\%, then the time of day and week should be taken into account, along with sufficient repetitions and, ideally, re-provisioning of the benchmarking infrastructure.
This is especially relevant when demonstrating slight performance improvements (e.g., in research paper) or attempting to detect early performance regressions.
If, however, only substantial performance differences are of interest as with most of our stream processing benchmarking at Dynatrace, then performance variability and precise timing become far less critical.
In such scenarios, we recommend not investing excessive time and cost into overly accurate measurements.

We provide all the collected data from our experiments as supplemental material~\cite{ReplicationPackage} for future analysis of cloud variability for the case of stream processing applications.
Future work could also further validate our findings or reveal their limitations. Our benchmarking setup allows easily repeating these experiments with other cloud providers, stream processing frameworks, virtual machine types, cloud regions, deployment sizes, or configuration of our benchmark.
Moreover, our study design is intended to serve as a template for conducting similar studies of cloud performance variability with other types of applications and software systems.

\begin{acks}
We would like to thank the Johannes Kepler University Linz and Dynatrace for co-funding this research.
\end{acks}

\bibliographystyle{ACM-Reference-Format}
\bibliography{references}


\begin{thebibliography}{37}


\ifx \showCODEN    \undefined \def \showCODEN     #1{\unskip}     \fi
\ifx \showDOI      \undefined \def \showDOI       #1{#1}\fi
\ifx \showISBNx    \undefined \def \showISBNx     #1{\unskip}     \fi
\ifx \showISBNxiii \undefined \def \showISBNxiii  #1{\unskip}     \fi
\ifx \showISSN     \undefined \def \showISSN      #1{\unskip}     \fi
\ifx \showLCCN     \undefined \def \showLCCN      #1{\unskip}     \fi
\ifx \shownote     \undefined \def \shownote      #1{#1}          \fi
\ifx \showarticletitle \undefined \def \showarticletitle #1{#1}   \fi
\ifx \showURL      \undefined \def \showURL       {\relax}        \fi
\providecommand\bibfield[2]{#2}
\providecommand\bibinfo[2]{#2}
\providecommand\natexlab[1]{#1}
\providecommand\showeprint[2][]{arXiv:#2}

\bibitem[Abedi and Brecht(2017)]%
        {Abedi2017}
\bibfield{author}{\bibinfo{person}{Ali Abedi} {and} \bibinfo{person}{Tim
  Brecht}.} \bibinfo{year}{2017}\natexlab{}.
\newblock \showarticletitle{Conducting Repeatable Experiments in Highly
  Variable Cloud Computing Environments}. In
  \bibinfo{booktitle}{\emph{Proceedings of the 8th ACM/SPEC on International
  Conference on Performance Engineering}} (L'Aquila, Italy)
  \emph{(\bibinfo{series}{ICPE '17})}. \bibinfo{publisher}{ACM},
  \bibinfo{pages}{287–292}.
\newblock
\showISBNx{9781450344043}
\urldef\tempurl%
\url{https://doi.org/10.1145/3030207.3030229}
\showDOI{\tempurl}


\bibitem[Baresi et~al\mbox{.}(2023)]%
        {Baresi2023}
\bibfield{author}{\bibinfo{person}{Luciano Baresi}, \bibinfo{person}{Tommaso
  Dolci}, \bibinfo{person}{Giovanni Quattrocchi}, {and}
  \bibinfo{person}{Nicholas Rasi}.} \bibinfo{year}{2023}\natexlab{}.
\newblock \showarticletitle{A multi-faceted analysis of the performance
  variability of virtual machines}.
\newblock \bibinfo{journal}{\emph{Software: Practice and Experience}}
  \bibinfo{volume}{53}, \bibinfo{number}{11} (\bibinfo{year}{2023}),
  \bibinfo{pages}{2067--2091}.
\newblock
\urldef\tempurl%
\url{https://doi.org/10.1002/spe.3244}
\showDOI{\tempurl}


\bibitem[Bermbach et~al\mbox{.}(2017)]%
        {Bermbach2017}
\bibfield{author}{\bibinfo{person}{David Bermbach}, \bibinfo{person}{Erik
  Wittern}, {and} \bibinfo{person}{Stefan Tai}.}
  \bibinfo{year}{2017}\natexlab{}.
\newblock \bibinfo{booktitle}{\emph{Cloud Service Benchmarking}
  (\bibinfo{edition}{1st} ed.)}.
\newblock \bibinfo{publisher}{Springer}.
\newblock
\urldef\tempurl%
\url{https://doi.org/10.1007/978-3-319-55483-9}
\showDOI{\tempurl}


\bibitem[Borysenkov et~al\mbox{.}(2025)]%
        {SANER2025}
\bibfield{author}{\bibinfo{person}{Dmytro Borysenkov}, \bibinfo{person}{Adriano
  Vogel}, \bibinfo{person}{Sören Henning}, {and} \bibinfo{person}{Esteban
  Perez-Wohlfeil}.} \bibinfo{year}{2025}\natexlab{}.
\newblock \showarticletitle{Analyzing Logs of Large-Scale Software Systems
  using Time Curves Visualization}. In \bibinfo{booktitle}{\emph{2025 IEEE
  International Conference on Software Analysis, Evolution and Reengineering
  (SANER)}}. \bibinfo{publisher}{IEEE}.
\newblock
\urldef\tempurl%
\url{https://doi.org/10.1109/SANER64311.2025.00038}
\showDOI{\tempurl}
\newblock
\shownote{In press}.


\bibitem[Bulej et~al\mbox{.}(2017)]%
        {Bulej2017}
\bibfield{author}{\bibinfo{person}{Lubom\'{\i}r Bulej},
  \bibinfo{person}{Vojtech Hork\'{y}}, {and} \bibinfo{person}{Petr T\r{u}ma}.}
  \bibinfo{year}{2017}\natexlab{}.
\newblock \showarticletitle{Do We Teach Useful Statistics for Performance
  Evaluation?}. In \bibinfo{booktitle}{\emph{Proceedings of the 8th ACM/SPEC on
  International Conference on Performance Engineering Companion}} (L'Aquila,
  Italy) \emph{(\bibinfo{series}{ICPE '17 Companion})}.
  \bibinfo{publisher}{ACM}, \bibinfo{pages}{185–189}.
\newblock
\showISBNx{9781450348997}
\urldef\tempurl%
\url{https://doi.org/10.1145/3053600.3053638}
\showDOI{\tempurl}


\bibitem[Bulej et~al\mbox{.}(2020)]%
        {Bulej2020}
\bibfield{author}{\bibinfo{person}{Lubom\'{\i}r Bulej},
  \bibinfo{person}{Vojt\v{e}ch Hork\'{y}}, \bibinfo{person}{Petr Tuma},
  \bibinfo{person}{Fran\c{c}ois Farquet}, {and} \bibinfo{person}{Aleksandar
  Prokopec}.} \bibinfo{year}{2020}\natexlab{}.
\newblock \showarticletitle{Duet Benchmarking: Improving Measurement Accuracy
  in the Cloud}. In \bibinfo{booktitle}{\emph{Proceedings of the ACM/SPEC
  International Conference on Performance Engineering}} (Edmonton AB, Canada)
  \emph{(\bibinfo{series}{ICPE '20})}. \bibinfo{publisher}{ACM},
  \bibinfo{pages}{100–107}.
\newblock
\showISBNx{9781450369916}
\urldef\tempurl%
\url{https://doi.org/10.1145/3358960.3379132}
\showDOI{\tempurl}


\bibitem[Eismann et~al\mbox{.}(2022)]%
        {Eismann2022}
\bibfield{author}{\bibinfo{person}{Simon Eismann}, \bibinfo{person}{Diego~Elias
  Costa}, \bibinfo{person}{Lizhi Liao}, \bibinfo{person}{Cor-Paul Bezemer},
  \bibinfo{person}{Weiyi Shang}, \bibinfo{person}{Andr\'{e} van Hoorn}, {and}
  \bibinfo{person}{Samuel Kounev}.} \bibinfo{year}{2022}\natexlab{}.
\newblock \showarticletitle{A case study on the stability of performance tests
  for serverless applications}.
\newblock \bibinfo{journal}{\emph{J. Syst. Softw.}} \bibinfo{volume}{189},
  \bibinfo{number}{C} (\bibinfo{date}{July} \bibinfo{year}{2022}),
  \bibinfo{numpages}{14}~pages.
\newblock
\showISSN{0164-1212}
\urldef\tempurl%
\url{https://doi.org/10.1016/j.jss.2022.111294}
\showDOI{\tempurl}


\bibitem[Ertl(2024)]%
        {Ertl2024a}
\bibfield{author}{\bibinfo{person}{Otmar Ertl}.}
  \bibinfo{year}{2024}\natexlab{}.
\newblock \showarticletitle{{UltraLogLog}: A Practical and More Space-Efficient
  Alternative to {HyperLogLog} for Approximate Distinct Counting}.
\newblock \bibinfo{journal}{\emph{Proc. VLDB Endow.}} \bibinfo{volume}{17},
  \bibinfo{number}{7} (\bibinfo{date}{May} \bibinfo{year}{2024}),
  \bibinfo{pages}{1655–1668}.
\newblock
\showISSN{2150-8097}
\urldef\tempurl%
\url{https://doi.org/10.14778/3654621.3654632}
\showDOI{\tempurl}


\bibitem[Folkerts et~al\mbox{.}(2013)]%
        {Folkerts2013}
\bibfield{author}{\bibinfo{person}{Enno Folkerts}, \bibinfo{person}{Alexander
  Alexandrov}, \bibinfo{person}{Kai Sachs}, \bibinfo{person}{Alexandru Iosup},
  \bibinfo{person}{Volker Markl}, {and} \bibinfo{person}{Cafer Tosun}.}
  \bibinfo{year}{2013}\natexlab{}.
\newblock \showarticletitle{Benchmarking in the Cloud: What It Should, Can, and
  Cannot Be}. In \bibinfo{booktitle}{\emph{Selected Topics in Performance
  Evaluation and Benchmarking}}, \bibfield{editor}{\bibinfo{person}{Raghunath
  Nambiar} {and} \bibinfo{person}{Meikel Poess}} (Eds.).
  \bibinfo{publisher}{Springer}, \bibinfo{address}{Berlin, Heidelberg},
  \bibinfo{pages}{173--188}.
\newblock
\urldef\tempurl%
\url{https://doi.org/10.1007/978-3-642-36727-4_12}
\showDOI{\tempurl}


\bibitem[Gkatzikis and Koutsopoulos(2013)]%
        {gkatzikis2013}
\bibfield{author}{\bibinfo{person}{Lazaros Gkatzikis} {and}
  \bibinfo{person}{Iordanis Koutsopoulos}.} \bibinfo{year}{2013}\natexlab{}.
\newblock \showarticletitle{Migrate or not? Exploiting dynamic task migration
  in mobile cloud computing systems}.
\newblock \bibinfo{journal}{\emph{IEEE Wireless Communications}}
  \bibinfo{volume}{20}, \bibinfo{number}{3} (\bibinfo{year}{2013}),
  \bibinfo{pages}{24--32}.
\newblock


\bibitem[Hasselbring(2021)]%
        {Hasselbring2021}
\bibfield{author}{\bibinfo{person}{Wilhelm Hasselbring}.}
  \bibinfo{year}{2021}\natexlab{}.
\newblock \showarticletitle{Benchmarking as Empirical Standard in Software
  Engineering Research}. In \bibinfo{booktitle}{\emph{Evaluation and Assessment
  in Software Engineering}} \emph{(\bibinfo{series}{EASE '21})}.
  \bibinfo{publisher}{ACM}, \bibinfo{pages}{457--462}.
\newblock
\urldef\tempurl%
\url{https://doi.org/10.1145/3463274.3463361}
\showDOI{\tempurl}


\bibitem[He et~al\mbox{.}(2021)]%
        {He2021}
\bibfield{author}{\bibinfo{person}{Sen He}, \bibinfo{person}{Tianyi Liu},
  \bibinfo{person}{Palden Lama}, \bibinfo{person}{Jaewoo Lee},
  \bibinfo{person}{In~Kee Kim}, {and} \bibinfo{person}{Wei Wang}.}
  \bibinfo{year}{2021}\natexlab{}.
\newblock \showarticletitle{Performance Testing for Cloud Computing with
  Dependent Data Bootstrapping}. In \bibinfo{booktitle}{\emph{2021 36th
  IEEE/ACM International Conference on Automated Software Engineering (ASE)}}.
  \bibinfo{publisher}{IEEE}, \bibinfo{pages}{666--678}.
\newblock
\urldef\tempurl%
\url{https://doi.org/10.1109/ASE51524.2021.9678687}
\showDOI{\tempurl}


\bibitem[He et~al\mbox{.}(2019)]%
        {He2019}
\bibfield{author}{\bibinfo{person}{Sen He}, \bibinfo{person}{Glenna Manns},
  \bibinfo{person}{John Saunders}, \bibinfo{person}{Wei Wang},
  \bibinfo{person}{Lori Pollock}, {and} \bibinfo{person}{Mary~Lou Soffa}.}
  \bibinfo{year}{2019}\natexlab{}.
\newblock \showarticletitle{A Statistics-Based Performance Testing Methodology
  for Cloud Applications}. In \bibinfo{booktitle}{\emph{Proceedings of the 2019
  27th ACM Joint Meeting on European Software Engineering Conference and
  Symposium on the Foundations of Software Engineering}} (Tallinn, Estonia)
  \emph{(\bibinfo{series}{ESEC/FSE 2019})}. \bibinfo{publisher}{ACM},
  \bibinfo{pages}{188–199}.
\newblock
\showISBNx{9781450355728}
\urldef\tempurl%
\url{https://doi.org/10.1145/3338906.3338912}
\showDOI{\tempurl}


\bibitem[Henning and Hasselbring(2022)]%
        {EMSE2022}
\bibfield{author}{\bibinfo{person}{S\"{o}ren Henning} {and}
  \bibinfo{person}{Wilhelm Hasselbring}.} \bibinfo{year}{2022}\natexlab{}.
\newblock \showarticletitle{A Configurable Method for Benchmarking Scalability
  of Cloud-Native Applications}.
\newblock \bibinfo{journal}{\emph{Empirical Software Engineering}}
  \bibinfo{volume}{27}, \bibinfo{number}{6} (\bibinfo{date}{Aug.}
  \bibinfo{year}{2022}).
\newblock
\urldef\tempurl%
\url{https://doi.org/10.1007/s10664-022-10162-1}
\showDOI{\tempurl}


\bibitem[Henning and Hasselbring(2024)]%
        {JSS2024}
\bibfield{author}{\bibinfo{person}{S{\"o}ren Henning} {and}
  \bibinfo{person}{Wilhelm Hasselbring}.} \bibinfo{year}{2024}\natexlab{}.
\newblock \showarticletitle{Benchmarking scalability of stream processing
  frameworks deployed as microservices in the cloud}.
\newblock \bibinfo{journal}{\emph{Journal of Systems and Software}}
  \bibinfo{volume}{208} (\bibinfo{year}{2024}), \bibinfo{pages}{111879}.
\newblock
\urldef\tempurl%
\url{https://doi.org/10.1016/j.jss.2023.111879}
\showDOI{\tempurl}


\bibitem[Henning et~al\mbox{.}(2024)]%
        {ICPE2024}
\bibfield{author}{\bibinfo{person}{S\"{o}ren Henning}, \bibinfo{person}{Adriano
  Vogel}, \bibinfo{person}{Michael Leichtfried}, \bibinfo{person}{Otmar Ertl},
  {and} \bibinfo{person}{Rick Rabiser}.} \bibinfo{year}{2024}\natexlab{}.
\newblock \showarticletitle{ShuffleBench: A Benchmark for Large-Scale Data
  Shuffling Operations with Distributed Stream Processing Frameworks}. In
  \bibinfo{booktitle}{\emph{Proceedings of the 15th ACM/SPEC International
  Conference on Performance Engineering}} (London, United Kingdom)
  \emph{(\bibinfo{series}{ICPE '24})}. \bibinfo{publisher}{ACM},
  \bibinfo{pages}{2–13}.
\newblock
\showISBNx{9798400704444}
\urldef\tempurl%
\url{https://doi.org/10.1145/3629526.3645036}
\showDOI{\tempurl}


\bibitem[Henning et~al\mbox{.}(2025)]%
        {ReplicationPackage}
\bibfield{author}{\bibinfo{person}{S\"{o}ren Henning}, \bibinfo{person}{Adriano
  Vogel}, \bibinfo{person}{Esteban Perez-Wohlfeil}, \bibinfo{person}{Otmar
  Ertl}, {and} \bibinfo{person}{Rick Rabiser}.}
  \bibinfo{year}{2025}\natexlab{}.
\newblock \bibinfo{booktitle}{\emph{Replication Package for: When Should I Run
  My Application Benchmark?: Studying Cloud Performance Variability for the
  Case of Stream Processing Applications}}.
\newblock
\urldef\tempurl%
\url{https://doi.org/10.5281/zenodo.14617187}
\showDOI{\tempurl}


\bibitem[Hesse et~al\mbox{.}(2021)]%
        {Hesse2021}
\bibfield{author}{\bibinfo{person}{Guenter Hesse}, \bibinfo{person}{Christoph
  Matthies}, \bibinfo{person}{Michael Perscheid}, \bibinfo{person}{Matthias
  Uflacker}, {and} \bibinfo{person}{Hasso Plattner}.}
  \bibinfo{year}{2021}\natexlab{}.
\newblock \showarticletitle{{ESPBench}: The Enterprise Stream Processing
  Benchmark}. In \bibinfo{booktitle}{\emph{Proceedings of the ACM/SPEC
  International Conference on Performance Engineering}}
  \emph{(\bibinfo{series}{ICPE '21})}. \bibinfo{publisher}{ACM},
  \bibinfo{pages}{201--212}.
\newblock
\showISBNx{9781450381949}
\urldef\tempurl%
\url{https://doi.org/10.1145/3427921.3450242}
\showDOI{\tempurl}


\bibitem[Iosup et~al\mbox{.}(2011)]%
        {Iosup2011}
\bibfield{author}{\bibinfo{person}{Alexandru Iosup}, \bibinfo{person}{Nezih
  Yigitbasi}, {and} \bibinfo{person}{Dick Epema}.}
  \bibinfo{year}{2011}\natexlab{}.
\newblock \showarticletitle{On the Performance Variability of Production Cloud
  Services}. In \bibinfo{booktitle}{\emph{2011 11th IEEE/ACM International
  Symposium on Cluster, Cloud and Grid Computing}}. \bibinfo{publisher}{IEEE},
  \bibinfo{pages}{104--113}.
\newblock
\urldef\tempurl%
\url{https://doi.org/10.1109/CCGrid.2011.22}
\showDOI{\tempurl}


\bibitem[Kounev et~al\mbox{.}(2020)]%
        {Kounev2020}
\bibfield{author}{\bibinfo{person}{Samuel Kounev},
  \bibinfo{person}{Klaus-Dieter Lange}, {and} \bibinfo{person}{J\'{o}akim von
  Kistowski}.} \bibinfo{year}{2020}\natexlab{}.
\newblock \bibinfo{booktitle}{\emph{Systems Benchmarking: For Scientists and
  Engineers} (\bibinfo{edition}{1st} ed.)}.
\newblock \bibinfo{publisher}{Springer}.
\newblock
\urldef\tempurl%
\url{https://doi.org/10.1007/978-3-030-41705-5}
\showDOI{\tempurl}


\bibitem[Laaber et~al\mbox{.}(2019)]%
        {Laaber2019}
\bibfield{author}{\bibinfo{person}{Christoph Laaber}, \bibinfo{person}{Joel
  Scheuner}, {and} \bibinfo{person}{Philipp Leitner}.}
  \bibinfo{year}{2019}\natexlab{}.
\newblock \showarticletitle{Software Microbenchmarking in the Cloud. How Bad is
  It Really?}
\newblock \bibinfo{journal}{\emph{Empirical Softw. Engg.}}
  \bibinfo{volume}{24}, \bibinfo{number}{4} (\bibinfo{date}{Aug.}
  \bibinfo{year}{2019}), \bibinfo{pages}{2469–2508}.
\newblock
\urldef\tempurl%
\url{https://doi.org/10.1007/s10664-019-09681-1}
\showDOI{\tempurl}


\bibitem[Leitner and Cito(2016)]%
        {Leitner2016}
\bibfield{author}{\bibinfo{person}{Philipp Leitner} {and}
  \bibinfo{person}{J\"{u}rgen Cito}.} \bibinfo{year}{2016}\natexlab{}.
\newblock \showarticletitle{Patterns in the Chaos—A Study of Performance
  Variation and Predictability in Public IaaS Clouds}.
\newblock \bibinfo{journal}{\emph{ACM Trans. Internet Technol.}}
  \bibinfo{volume}{16}, \bibinfo{number}{3}, Article \bibinfo{articleno}{15}
  (\bibinfo{date}{April} \bibinfo{year}{2016}), \bibinfo{numpages}{23}~pages.
\newblock
\urldef\tempurl%
\url{https://doi.org/10.1145/2885497}
\showDOI{\tempurl}


\bibitem[Maricq et~al\mbox{.}(2018)]%
        {Maricq2018}
\bibfield{author}{\bibinfo{person}{Aleksander Maricq}, \bibinfo{person}{Dmitry
  Duplyakin}, \bibinfo{person}{Ivo Jimenez}, \bibinfo{person}{Carlos Maltzahn},
  \bibinfo{person}{Ryan Stutsman}, {and} \bibinfo{person}{Robert Ricci}.}
  \bibinfo{year}{2018}\natexlab{}.
\newblock \showarticletitle{Taming Performance Variability}. In
  \bibinfo{booktitle}{\emph{Proceedings of the 13th USENIX Conference on
  Operating Systems Design and Implementation}} (Carlsbad, CA, USA)
  \emph{(\bibinfo{series}{OSDI'18})}. \bibinfo{publisher}{USENIX Association},
  \bibinfo{pages}{409--425}.
\newblock


\bibitem[Netto et~al\mbox{.}(2018)]%
        {Netto2018}
\bibfield{author}{\bibinfo{person}{Marco A.~S. Netto},
  \bibinfo{person}{Rodrigo~N. Calheiros}, \bibinfo{person}{Eduardo~R.
  Rodrigues}, \bibinfo{person}{Renato L.~F. Cunha}, {and}
  \bibinfo{person}{Rajkumar Buyya}.} \bibinfo{year}{2018}\natexlab{}.
\newblock \showarticletitle{{HPC} Cloud for Scientific and Business
  Applications: Taxonomy, Vision, and Research Challenges}.
\newblock \bibinfo{journal}{\emph{Comput. Surveys}} \bibinfo{volume}{51},
  \bibinfo{number}{1}, Article \bibinfo{articleno}{8} (\bibinfo{date}{Jan.}
  \bibinfo{year}{2018}), \bibinfo{numpages}{29}~pages.
\newblock
\showISSN{0360-0300}
\urldef\tempurl%
\url{https://doi.org/10.1145/3150224}
\showDOI{\tempurl}


\bibitem[Papadopoulos et~al\mbox{.}(2021)]%
        {Papadopoulos2021}
\bibfield{author}{\bibinfo{person}{Alessandro~Vittorio Papadopoulos},
  \bibinfo{person}{Laurens Versluis}, \bibinfo{person}{André Bauer},
  \bibinfo{person}{Nikolas Herbst}, \bibinfo{person}{Jóakim~von Kistowski},
  \bibinfo{person}{Ahmed Ali-Eldin}, \bibinfo{person}{Cristina~L. Abad},
  \bibinfo{person}{José~Nelson Amaral}, \bibinfo{person}{Petr Tůma}, {and}
  \bibinfo{person}{Alexandru Iosup}.} \bibinfo{year}{2021}\natexlab{}.
\newblock \showarticletitle{Methodological Principles for Reproducible
  Performance Evaluation in Cloud Computing}.
\newblock \bibinfo{journal}{\emph{IEEE Transactions on Software Engineering}}
  \bibinfo{volume}{47}, \bibinfo{number}{8} (\bibinfo{year}{2021}),
  \bibinfo{pages}{1528--1543}.
\newblock
\urldef\tempurl%
\url{https://doi.org/10.1109/TSE.2019.2927908}
\showDOI{\tempurl}


\bibitem[Reichinger et~al\mbox{.}(2024)]%
        {Reichinger2024}
\bibfield{author}{\bibinfo{person}{Julian Reichinger}, \bibinfo{person}{Thomas
  Krismayer}, {and} \bibinfo{person}{Jan Rellermeyer}.}
  \bibinfo{year}{2024}\natexlab{}.
\newblock \bibinfo{title}{{COPR} -- Efficient, large-scale log storage and
  retrieval}.
\newblock
\newblock
\urldef\tempurl%
\url{https://doi.org/10.48550/arXiv.2402.18355}
\showDOI{\tempurl}
\showeprint[arxiv]{2402.18355}~[cs.IR]


\bibitem[Sax et~al\mbox{.}(2018)]%
        {Sax2018}
\bibfield{author}{\bibinfo{person}{Matthias~J. Sax}, \bibinfo{person}{Guozhang
  Wang}, \bibinfo{person}{Matthias Weidlich}, {and}
  \bibinfo{person}{Johann-Christoph Freytag}.} \bibinfo{year}{2018}\natexlab{}.
\newblock \showarticletitle{Streams and Tables: Two Sides of the Same Coin}. In
  \bibinfo{booktitle}{\emph{Proceedings of the International Workshop on
  Real-Time Business Intelligence and Analytics}} \emph{(\bibinfo{series}{BIRTE
  '18})}. \bibinfo{publisher}{ACM}, \bibinfo{numpages}{10}~pages.
\newblock
\showISBNx{9781450366076}
\urldef\tempurl%
\url{https://doi.org/10.1145/3242153.3242155}
\showDOI{\tempurl}


\bibitem[Schad et~al\mbox{.}(2010)]%
        {Schad2010}
\bibfield{author}{\bibinfo{person}{J\"{o}rg Schad}, \bibinfo{person}{Jens
  Dittrich}, {and} \bibinfo{person}{Jorge-Arnulfo Quian\'{e}-Ruiz}.}
  \bibinfo{year}{2010}\natexlab{}.
\newblock \showarticletitle{Runtime measurements in the cloud: observing,
  analyzing, and reducing variance}.
\newblock \bibinfo{journal}{\emph{Proc. VLDB Endow.}} \bibinfo{volume}{3},
  \bibinfo{number}{1–2} (\bibinfo{date}{Sept.} \bibinfo{year}{2010}),
  \bibinfo{pages}{460–471}.
\newblock
\showISSN{2150-8097}
\urldef\tempurl%
\url{https://doi.org/10.14778/1920841.1920902}
\showDOI{\tempurl}


\bibitem[Schirmer et~al\mbox{.}(2023)]%
        {Schirmer2023}
\bibfield{author}{\bibinfo{person}{Trever Schirmer}, \bibinfo{person}{Nils
  Japke}, \bibinfo{person}{Sofia Greten}, \bibinfo{person}{Tobias Pfandzelter},
  {and} \bibinfo{person}{David Bermbach}.} \bibinfo{year}{2023}\natexlab{}.
\newblock \showarticletitle{The Night Shift: Understanding Performance
  Variability of Cloud Serverless Platforms}. In
  \bibinfo{booktitle}{\emph{Proceedings of the 1st Workshop on SErverless
  Systems, Applications and MEthodologies}} (Rome, Italy)
  \emph{(\bibinfo{series}{SESAME '23})}. \bibinfo{publisher}{ACM},
  \bibinfo{pages}{27–33}.
\newblock
\showISBNx{9798400701856}
\urldef\tempurl%
\url{https://doi.org/10.1145/3592533.3592808}
\showDOI{\tempurl}


\bibitem[Stol and Fitzgerald(2018)]%
        {Stol2018}
\bibfield{author}{\bibinfo{person}{Klaas-Jan Stol} {and} \bibinfo{person}{Brian
  Fitzgerald}.} \bibinfo{year}{2018}\natexlab{}.
\newblock \showarticletitle{The ABC of Software Engineering Research}.
\newblock \bibinfo{journal}{\emph{ACM Trans. Softw. Eng. Methodol.}}
  \bibinfo{volume}{27}, \bibinfo{number}{3}, Article \bibinfo{articleno}{11}
  (\bibinfo{date}{Sept.} \bibinfo{year}{2018}), \bibinfo{numpages}{51}~pages.
\newblock
\showISSN{1049-331X}
\urldef\tempurl%
\url{https://doi.org/10.1145/3241743}
\showDOI{\tempurl}


\bibitem[Uta et~al\mbox{.}(2020)]%
        {Uta2020}
\bibfield{author}{\bibinfo{person}{Alexandru Uta}, \bibinfo{person}{Alexandru
  Custura}, \bibinfo{person}{Dmitry Duplyakin}, \bibinfo{person}{Ivo Jimenez},
  \bibinfo{person}{Jan Rellermeyer}, \bibinfo{person}{Carlos Maltzahn},
  \bibinfo{person}{Robert Ricci}, {and} \bibinfo{person}{Alexandru Iosup}.}
  \bibinfo{year}{2020}\natexlab{}.
\newblock \showarticletitle{Is big data performance reproducible in modern
  cloud networks?}. In \bibinfo{booktitle}{\emph{Proceedings of the 17th Usenix
  Conference on Networked Systems Design and Implementation}} (Santa Clara, CA,
  USA) \emph{(\bibinfo{series}{NSDI'20})}. \bibinfo{publisher}{USENIX
  Association}, \bibinfo{pages}{513–528}.
\newblock
\showISBNx{9781939133137}


\bibitem[Vogel et~al\mbox{.}(2023)]%
        {SEAA2023}
\bibfield{author}{\bibinfo{person}{Adriano Vogel}, \bibinfo{person}{S\"{o}ren
  Henning}, \bibinfo{person}{Otmar Ertl}, {and} \bibinfo{person}{Rick
  Rabiser}.} \bibinfo{year}{2023}\natexlab{}.
\newblock \showarticletitle{A systematic mapping of performance in distributed
  stream processing systems}. In \bibinfo{booktitle}{\emph{Euromicro Conference
  on Software Engineering and Advanced Applications}}.
  \bibinfo{publisher}{IEEE}.
\newblock
\urldef\tempurl%
\url{https://doi.org/10.1109/SEAA60479.2023.00052}
\showDOI{\tempurl}


\bibitem[Vogel et~al\mbox{.}(2024)]%
        {DEBS2024}
\bibfield{author}{\bibinfo{person}{Adriano Vogel}, \bibinfo{person}{S\"{o}ren
  Henning}, \bibinfo{person}{Esteban Perez-Wohlfeil}, \bibinfo{person}{Otmar
  Ertl}, {and} \bibinfo{person}{Rick Rabiser}.}
  \bibinfo{year}{2024}\natexlab{}.
\newblock \showarticletitle{A Comprehensive Benchmarking Analysis of Fault
  Recovery in Stream Processing Frameworks}. In
  \bibinfo{booktitle}{\emph{Proceedings of the 18th ACM International
  Conference on Distributed and Event-Based Systems}} (Villeurbanne, France)
  \emph{(\bibinfo{series}{DEBS '24})}. \bibinfo{publisher}{ACM},
  \bibinfo{pages}{171–182}.
\newblock
\showISBNx{9798400704437}
\urldef\tempurl%
\url{https://doi.org/10.1145/3629104.3666040}
\showDOI{\tempurl}


\bibitem[Wang et~al\mbox{.}(2021)]%
        {Wang2021}
\bibfield{author}{\bibinfo{person}{Guozhang Wang}, \bibinfo{person}{Lei Chen},
  \bibinfo{person}{Ayusman Dikshit}, \bibinfo{person}{Jason Gustafson},
  \bibinfo{person}{Boyang Chen}, \bibinfo{person}{Matthias~J. Sax},
  \bibinfo{person}{John Roesler}, \bibinfo{person}{Sophie Blee-Goldman},
  \bibinfo{person}{Bruno Cadonna}, \bibinfo{person}{Apurva Mehta},
  \bibinfo{person}{Varun Madan}, {and} \bibinfo{person}{Jun Rao}.}
  \bibinfo{year}{2021}\natexlab{}.
\newblock \showarticletitle{Consistency and Completeness: Rethinking
  Distributed Stream Processing in {Apache Kafka}}. In
  \bibinfo{booktitle}{\emph{Proceedings of the 2021 International Conference on
  Management of Data}} \emph{(\bibinfo{series}{SIGMOD/PODS '21})}.
  \bibinfo{publisher}{ACM}, \bibinfo{pages}{2602–2613}.
\newblock
\urldef\tempurl%
\url{https://doi.org/10.1145/3448016.3457556}
\showDOI{\tempurl}


\bibitem[Wen et~al\mbox{.}(2025)]%
        {Wen2025}
\bibfield{author}{\bibinfo{person}{Jinfeng Wen}, \bibinfo{person}{Zhenpeng
  Chen}, \bibinfo{person}{Federica Sarro}, {and} \bibinfo{person}{Shangguang
  Wang}.} \bibinfo{year}{2025}\natexlab{}.
\newblock \showarticletitle{Unveiling Overlooked Performance Variance in
  Serverless Computing}.
\newblock \bibinfo{journal}{\emph{Empirical Softw. Engg.}}
  \bibinfo{volume}{30}, \bibinfo{number}{2} (\bibinfo{date}{Aug.}
  \bibinfo{year}{2025}).
\newblock
\urldef\tempurl%
\url{https://doi.org/10.1007/s10664-025-10615-3}
\showDOI{\tempurl}


\bibitem[Wohlin and Rainer(2022)]%
        {Wohlin2022}
\bibfield{author}{\bibinfo{person}{Claes Wohlin} {and} \bibinfo{person}{Austen
  Rainer}.} \bibinfo{year}{2022}\natexlab{}.
\newblock \showarticletitle{Is it a case study?--A critical analysis and
  guidance}.
\newblock \bibinfo{journal}{\emph{Journal of Systems and Software}}
  \bibinfo{volume}{192} (\bibinfo{year}{2022}), \bibinfo{pages}{111395}.
\newblock
\showISSN{0164-1212}
\urldef\tempurl%
\url{https://doi.org/10.1016/j.jss.2022.111395}
\showDOI{\tempurl}


\bibitem[Zhao et~al\mbox{.}(2021)]%
        {Zhao2021}
\bibfield{author}{\bibinfo{person}{Yuxuan Zhao}, \bibinfo{person}{Dmitry
  Duplyakin}, \bibinfo{person}{Robert Ricci}, {and} \bibinfo{person}{Alexandru
  Uta}.} \bibinfo{year}{2021}\natexlab{}.
\newblock \showarticletitle{Cloud Performance Variability Prediction}. In
  \bibinfo{booktitle}{\emph{Companion of the ACM/SPEC International Conference
  on Performance Engineering}} (Virtual Event, France)
  \emph{(\bibinfo{series}{ICPE '21})}. \bibinfo{publisher}{ACM},
  \bibinfo{pages}{35–40}.
\newblock
\showISBNx{9781450383318}
\urldef\tempurl%
\url{https://doi.org/10.1145/3447545.3451182}
\showDOI{\tempurl}


\end{thebibliography}

\end{document}